# Light-Induced Lattice Coherence and Emission Enhancement in PTM-Passivated CsSnI$_3$ Perovskites

*Thomas Y. Adams, Bruce Barrios, Michael Ziegenfus, Hui Cai, and Sayantani Ghosh\**

T. Y. Adams, B. Barrios, H. Cai, S. Ghosh
Department of Physics, University of California Merced, 95343, United States of America
E-mail: sghosh@ucmerced.edu

M. Zeigenfuss
Radford University, Radford, VA 24142, United States of America

Metal halide perovskites continue to lead in optoelectronic applications, but the toxicity of lead has driven efforts to identify environmentally benign alternatives. Cesium tin iodide (CsSnI$_3$) is one such, with a direct bandgap and near-infrared emission, though its performance is limited by instability. We show that phthalimide (PTM) passivation during single crystal growth enhances optical output and ambient stability. Under continuous excitation, PTM-passivated microscale crystals show up to a nearly one order of magnitude increase in photoluminescence (PL) quantum yield, accompanied by reversible sharpening of a low-frequency Raman mode associated with Cs$^+$ rattling. This reveals dynamic, light-induced lattice reordering that passivates trap states and enhances radiative recombination. Mechanical grinding yields nanocrystals with redshifted, narrowed PL, consistent with a relaxed polymorph and reduced inhomogeneous broadening. Despite increased surface area, PTM remains effective in preserving near-infrared emission in nanocrystals as well. Power-dependent PL reveals distinct carrier dynamics: microcrystals show redshift due to bandgap renormalization, while nanocrystals show blueshift and elevated carrier temperatures (300–1900 K), consistent with hot-carrier recombination. Extended illumination reveals reversible optical changes, including PL modulation, reflecting dynamic light–matter interactions and evolving defect landscapes. These results identify PTM-passivated CsSnI$_3$ as an ideal platform for probing morphology-dependent carrier relaxation and light-induced vibrational coherence in lead-free perovskites.

# 1. Introduction

Metal halide perovskites (MHPs) with the general formula $ABX_3$ have garnered sustained interest for their exceptional optoelectronic properties, including high absorption coefficients, low exciton binding energies, and defect tolerance. The tunability of their composition through substitution at the A-site (e.g., $Cs^+$, $MA^+$, $FA^+$), B-site (e.g., $Pb^{2+}$, $Sn^{2+}$), and X-site ($I^-$, $Br^-$, $Cl^-$) has enabled performance breakthroughs in photovoltaics, light-emitting diodes, photodetectors, and other optoelectronic applications. However, the toxicity of Pb-based perovskites remains a critical barrier to large-scale deployment, prompting efforts to identify lead-free alternatives with comparable functionality.[1] Among these, tin-based halide perovskites have emerged as promising candidates. The divalent $Sn^{2+}$ ion offers a similar ionic radius and valency to $Pb^{2+}$ but enables materials with narrower bandgaps and stronger near-infrared (NIR) emission, which are ideal for applications such as in vivo bioimaging and NIR photovoltaics.[2-6] While $Sn^{2+}$ is not completely benign, quantitative studies show that Sn-based perovskite precursors and nanocrystals exhibit orders-of-magnitude lower cytotoxicity and aquatic toxicity than their Pb-containing analogues, and bulk Sn halide salts have negligible aqueous solubility, greatly reducing environmental exposure.[7] However, $Sn^{2+}$ is highly susceptible to oxidation, particularly under ambient conditions, leading to the formation of $Sn^{4+}$, p-type self-doping, and rapid degradation of optical properties. In the all-inorganic system $CsSnI_3$, this oxidation drives a phase transformation from the black, photoactive orthorhombic perovskite phase (~1.3 eV bandgap) to a yellow polymorph (~2.5 eV bandgap), limiting its utility in practical devices.[8, 9]

Efforts to stabilize Sn-based perovskites have included chemical doping, compositional engineering, and the use of defect-passivating additives.[10-13] Stabilization strategies for Sn-based perovskites fall broadly into three categories: additive/surface passivation, doping/compositional engineering, and encapsulation or interface protection.[14] Phthalimide (PTM), a small molecule containing electron-rich carbonyl and imide groups, has recently been shown to coordinate with $Sn^{2+}$, reducing surface oxidation and trap density in thin films.[15] PTM belongs to the class of additive/surface passivation, acting through targeted CO/NH coordination with $Sn^{2+}$ to suppress vacancy formation and oxidation without perturbing the bulk lattice. In contrast to dopants that modify electronic structure and encapsulation layers that physically block moisture and oxygen, PTM directly stabilizes the intrinsic $Sn^{2+}$ chemistry and remains compatible with both approaches. Studies to date have focused primarily on polycrystalline or film-based systems, leaving open questions about the intrinsic photophysical response of high-quality, PTM-stabilized CsSnI$_3$ single crystals. While chemical stabilization strategies have been extensively applied to Sn-based thin films, fundamental questions remain regarding how lattice dynamics and morphology modulate carrier recombination and photostability in single crystals. CsSnI$_3$'s soft lattice and strong carrier–phonon coupling also suggest the potential for dynamic lattice responses under illumination, but the extent to which such effects influence emission in passivated crystals has not been systematically explored. Moreover, the interplay between structural dimensionality and carrier thermalization remains poorly understood in lead-free perovskites, particularly in systems that permit direct comparisons between phonon-cooled and hot-carrier regimes within a unified chemical framework.

In this work, we present a comprehensive investigation of PTM-passivated CsSnI$_3$ microscale single crystals and their nanocrystalline derivatives, focusing on the interplay between lattice dynamics, trap state modulation, and carrier relaxation. We demonstrate that PTM not only stabilizes the orthorhombic phase under ambient conditions but also induces a clearly observable, reversible photoluminescence (PL) enhancement linked to coherent A-site Cs$^+$ vibrational modes. Using time-resolved Raman and power-dependent PL spectroscopy, we identify a photoinduced ordering process that passivates shallow traps and modulates carrier recombination. By reducing the microcrystals to the nanoscale, we uncover a transition from phonon-dominated relaxation to hot-carrier emission, confirmed by carrier temperature extraction from PL tail fitting. These findings reveal that CsSnI$_3$ supports morphology-tunable carrier dynamics, reversible structural reordering under light exposure, and dynamic defect healing, making it a versatile platform for probing and engineering light–matter interactions in lead-free perovskites.

## 2. Results and Discussion
### 2.1. Synthesis and structural characterization

CsSnI$_3$ microscale single crystals (MSCs) were synthesized by drop-casting a 0.267 M CsI/SnI$_2$ precursor solution onto glass substrates, followed by annealing at 370 °C. This annealing temperature was selected based on prior reports showing that CsSnI$_3$ annealed between 200–400 °C exhibits optimal optical quality at 370 °C, due to enhanced phase purity and reduced defect formation.[16] A confocal fluorescence image of the resulting as-synthesized crystals is shown in **Figure 1A**. Optical microscopy reveals two distinct morphologies: orthorhombic crystals with the black-phase structure (**Figure 1B**), and a second population of tetragonal crystals (**Figure 1C**). PL spectra comparing the two phases are shown in **Figure 1D**, where the orthorhombic phase exhibits near-infrared PL at 869 nm, while the tetragonal crystals show a slightly redshifted emission at 879 nm, attributed to increased lattice symmetry and reduced octahedral tilting.[17-19] These emissions stand in contrast to the 950 nm PL typically observed in $CsSnI_3$ thin films, a consequence of poor crystallinity, excess iodide, under-annealing, or the presence of Sn vacancies. The use of low precursor concentration and controlled annealing in this synthesis avoids such issues, enabling the isolation of phase-pure, structurally coherent MSCs suitable for probing intrinsic optical properties.

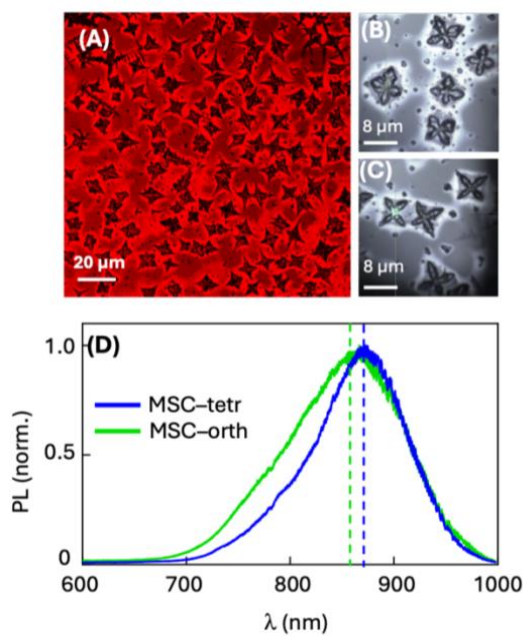

**Figure 1** (A) Confocal fluorescence microscopy image of ensemble of MSCs. Optical microscopy images of (B) orthorhombic and (C) tetragonal MSCs. (D) PL emission from individual MSCs of each phase. Dashed lines indicate the peaks of the spectra.

## 2.2. Optical Enhancement and Stability

To improve the optical stability and suppress non-radiative losses in $CsSnI_3$, PTM has been introduced in prior works as a surface passivator. PTM is a small organic molecule featuring conjugated carbonyl (C=O) and imide (–NH) functional groups, both of which possess lone pairs capable of coordinating with undercoordinated metal centers. In the context of $CsSnI_3$, these lone pairs form weak Lewis base interactions with surface $Sn^{2+}$ ions, passivating potential defect sites and simultaneously inhibiting oxidation to $Sn^{4+}$. In this way, PTM acts analogously to an antioxidant, stabilizing the $Sn^{2+}$ oxidation state and preventing the formation of shallow trap states that quench PL. Prior mechanistic studies[15] have established the role of PTM in suppressing oxidation to $Sn^{4+}$. Our work leverages this established chemistry and demonstrates that the inclusion of PTM in the precursor solution prior to annealing yields dramatic improvements in the optical quality and ambient stability of the resulting crystals. As shown in **Figure 2A**, PTM-passivated MSCs show stronger absorption and PL intensity than unpassivated samples. This is consistent with the suppression of Sn vacancies, which are known to act as non-radiative centers in Sn-based perovskites. X-ray diffraction (*Supplementary Figure S1*) shows that PTM does not incorporate into the crystal lattice: both PTM-treated and untreated crystals show peak positions characteristic of the orthorhombic phase, with no detectable shift in lattice parameters. However, PTM does influence the crystallization dynamics. As illustrated in **Figure 2B**, the average lateral size of crystals decreases from ~10 μm in the unpassivated MSCs to ~5 μm when PTM is added. This trend likely results from a reduction in the nucleation barrier, leading to earlier and more numerous nucleation events during growth. Additionally, PTM binding to specific crystal facets may inhibit growth along certain directions, further contributing to the observed size reduction.

Stability testing under ambient conditions further underscores the effectiveness of PTM. In unpassivated MSCs, PL intensity drops by an order of magnitude over 120 minutes of ambient exposure (**Figure 2C**). This decay is attributed to surface oxidation where $Sn^{2+}$ is gradually converted to $Sn^{4+}$, introducing deep trap states that quench radiative recombination. In contrast, PTM-passivated crystals show no measurable changes over 180 minutes (**Figure 2D**). These datasets inherently provide side-by-side stability comparisons under continuous illumination, revealing that PTM treatment slows PL degradation relative to unpassivated controls. We have not conducted thermal aging, thermal cycling, or device-level operational stability measurements, the ambient conditions have indirectly allowed us to verify stability over a relative humidity (RH) range of 42 – 57% (details in Experimental Methods Section).

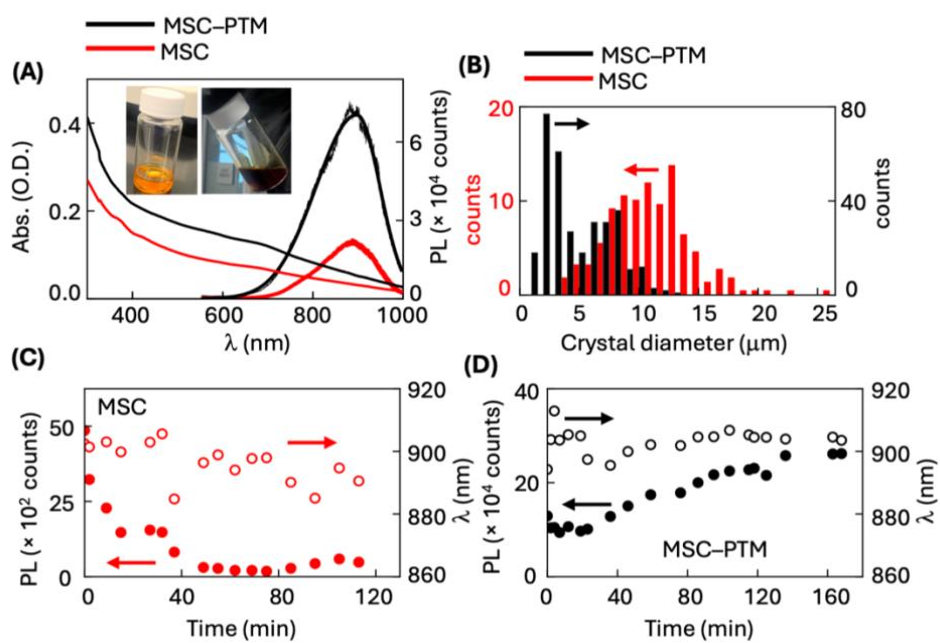

**Figure 2** (A) PL and absorption spectra of MSCs synthesized with and without PTM. *Insets*: solution mixtures of both kinds, with addition of PTM resulting in a darker tint. (B) MSC size distribution showing smaller crystal diameter on average on addition of PTM. PL intensity (solid circles) and wavelength (open circles) monitored over time for MSCs (C) without and (D) with PTM under identical ambient conditions (RH = 42%).

**2.3. Excitation-Dependent Emission Behavior in Single Crystals**

Continuing from the structural and passivation insights of **Figure 2**, we next examine the power- and time-dependent optical response of PTM-passivated $CsSnI_3$ single crystals under steady illumination. As shown in **Figure 3A**, emission wavelength exhibits three distinct regimes with increasing excitation power. At low powers the emission wavelength remains steady; in the intermediate regime, the emission redshifts from ~850 nm to 890 nm, and at high excitation powers, it saturates. The PL intensity increases rapidly at low powers while the emission wavelength remains steady (*Supplementary Figure S2*). As excitation power increases further, the PL intensity begins to saturate and a redshift in the emission emerges, consistent with many-body bandgap renormalization (BGR), in which high carrier densities screen Coulomb interactions, narrowing the bandgap. Fitting the PL emission energy $E(P)$ varying with incident power $P$ using $E(P) = E_O - \Delta E \log\left(P/P_O + 1\right)$ where $E_O$ is the initial band gap, $\Delta E$ the resulting redshift, and $P_O$ the initial power, yields BGR parameters of $E_O$ =1.48 eV, and $\Delta E$ = 0.07 eV that closely match observed behavior (*Supplementary Figure S3*). However, at higher excitation powers, the data begins to deviate from the BGR-only trend. This deviation may reflect the onset of Burstein–Moss (BM) effects, where band filling by photogenerated carriers leads to a blueshift that offsets the redshift from BGR.[20] The observed saturation in PL energy suggests that the system may be approaching a balance between these opposing effects. In this regime, the total spectral response represents a convolution of redshift from BGR and potential blueshift from hot carrier occupation of band-edge states, though no direct evidence of hot-carrier PL is seen, possibly due to efficient carrier cooling.

The absence of hot-carrier signatures in the PTM-passivated MSCs is notable, particularly given that $CsSnI_3$ thin films often exhibit power-dependent blueshifts consistent with hot carrier PL. In our case, several factors likely contribute to efficient carrier cooling that suppresses such effects. First, the high crystal quality and reduced defect density achieved through PTM passivation promote strong carrier–phonon coupling, enabling rapid thermalization via LO phonon emission and anharmonic phonon decay. Additionally, the large thermal mass and three-dimensional nature of the single crystals support effective heat dissipation through lattice conduction. Carrier diffusion also plays a role: in a clean crystalline system, photogenerated carriers excited near the surface can diffuse inward and cool before recombination, further limiting the persistence of non-equilibrium populations. Finally, the spectral evolution is dominated by BGR, leaving little opportunity for hot-carrier features to manifest.

## 2.4. Light-Induced Vibrational Reordering

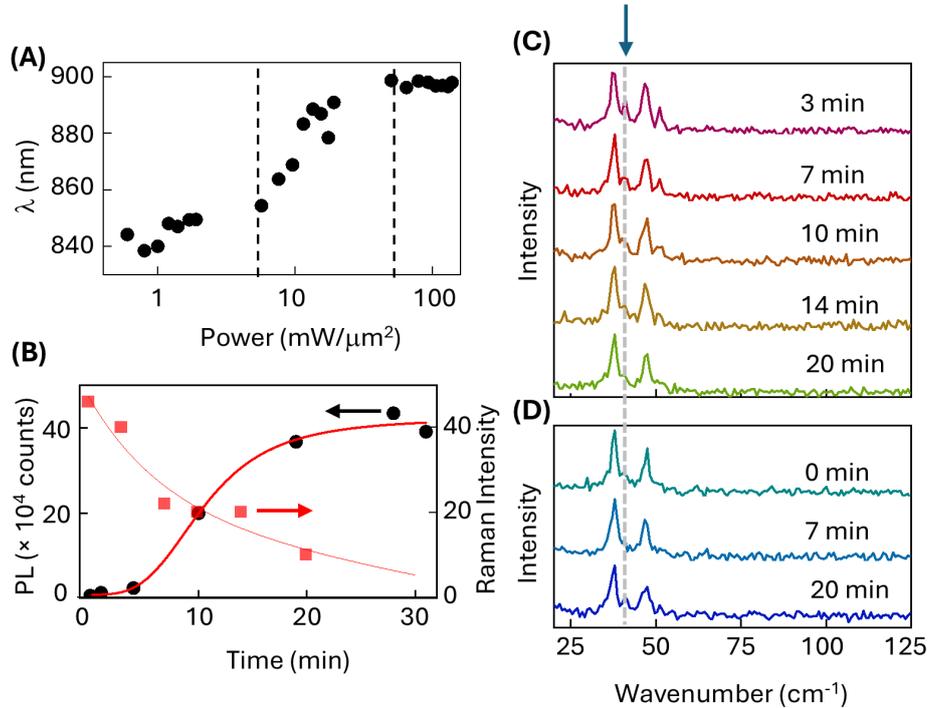

**Figure 3** (A) Emission wavelength demonstrating redshift with increasing excitation power. Dashed lines enclose the excitation power range over which the fastest red-shift occurs. (B) PL emission intensity (solid circles) increasing over time under continuous incident excitation, while intensity of the 40.8 cm$^{-1}$ Raman peak (solid squares) shows a simultaneous decrease. Lines are fits that indicate the PL rise time constant is 10 min, while the Raman signal decay constant is 7.8 min. (C) Raman spectra collected concurrently where the arrow indicates position of 40.8 cm$^{-1}$ peak. (D) Raman spectra collected once continuous excitation is ceased (RH = 41%).

To explore how continuous illumination affects the lattice, we tracked PL and Raman spectra under steady 532 nm excitation. As shown in **Figure 3B** and **3C**, the system evolves in two main stages. During the first few minutes, the PL intensity remains relatively stable, and the low-frequency Raman spectrum exhibits a sharp and distinct 40.8 cm$^{-1}$ mode, associated with A-site Cs$^+$ rattling.[21, 22] This mode is a vibrational signature of the orthorhombic phase and is highly sensitive to light-induced lattice relaxation and local symmetry. Notably, a rapid redshift in emission wavelength occurs early during this stage, after which the peak position stabilizes near 920 nm (*Supplementary Figure S4*). Between 5 and 20 minutes, as **Figure 3B** shows, PL intensity increases sharply, while the 40.8 cm$^{-1}$ mode fades, consistent with structural relaxation toward a more stable configuration. Beyond 20 minutes, the PL signal plateaus indicating that the lattice has reached a new quasi-equilibrium under illumination. In absolute terms, at this point the PTM-passivated crystals exhibit a PLQY of 11.2%, This magnitude is consistent with the known emissivity limits of Sn-based perovskites, whose PLQY is constrained by $Sn^{2+}$ oxidation, high intrinsic hole density, and Sn-vacancy–related nonradiative pathways.[23, 24] Reported $CsSnI_3$ PLQYs range from <1% in unoptimized nanocrystals[25] to ~10% in stabilized β-$CsSnI_3$ films. [26] By contrast, state-of-the-art Pb-based perovskite nanocrystals routinely achieve near-unity PLQY and $MAPbI_3$-family films can exhibit external PLQE above 50%,[27-30] underscoring the intrinsic emissivity gap between Sn- and Pb-halide perovskites.

This lattice evolution is reversible *(Supplementary Figure S5)*. Figure 3D shows that upon turning off the illumination, the Raman features recover to their original shape over ~20 minutes, indicating that the vibrational reordering is reversible and light-modulated. This dynamic behavior highlights the role of A-site lattice motion as a proxy for structural coherence, and its modulation by photoexcited carriers, and suggests that the observed vibrational dynamics do not stem from permanent chemical changes but rather from a photoresponsive modulation of the local lattice symmetry. A contribution from reversible redistribution of shallow point defects (e.g., iodine vacancies or hole-rich regions associated with Sn vacancies) cannot be ruled out; however, such processes typically broaden vibrational modes rather than sharpen them and therefore cannot account for the distinct 40.8 $cm^{-1}$ signature we observe. Illumination perturbs the Sn–I framework, likely expanding the octahedral cages and enabling increased mobility of the $Cs^+$ ions. This softening of the lattice facilitates the emergence of the rattling mode. As defects are passivated and the lattice reorganizes, the motion is damped. When the excitation is removed, the lattice relaxes and the dynamic disorder, and its associated vibrational signatures, are restored. Light-induced halide migration would be expected to broaden or quench Sn–I stretching modes and suppress NIR PL, which is opposite to the observed sharpening of the 40.8 cm$^{-1}$ mode and concurrent PL enhancement, further supporting a reversible lattice-reordering mechanism.

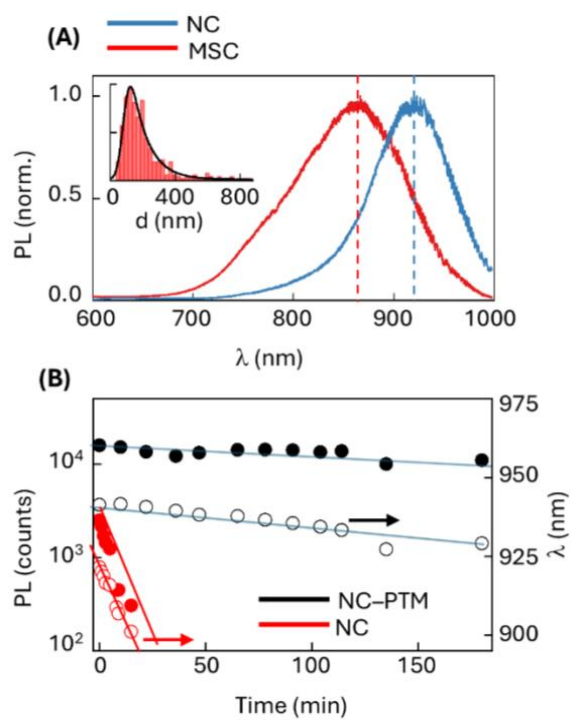

**Figure 4** (A) PL spectra comparing emissions of MSCs and NCs. *Inset*: size distribution of NCs. (B) PL intensity (solid circles) and wavelength (open circles) monitored over time for NC without and with PTM under identical ambient conditions (RH = 52%).

**2.5. Morphology-Dependent Optical and Thermal Behavior**

To investigate how morphology affects optical behavior, PTM-passivated MSCs were mechanically ground into nanocrystals (NCs) and compared with the parent material. While XRD confirmed the B-γ CsSnI$_3$ phase (*Supplementary Figure S6*),[31] As shown in **Figure 4A**, NCs exhibit a redshifted PL peak at 920 nm compared to 880 nm in the MSCs, along with spectral narrowing. This shift and narrowing suggest a transition to a lower bandgap, and more structurally coherent polymorph stabilized at the nanoscale. The average NC size is ~150 nm, a regime in which lattice reconstruction and local ordering become energetically favorable. PL measurements over time (**Figure 4B**) confirm that PTM remains effective in stabilizing the emission against ambient degradation, even with the increased surface area of the NCs.

Building on the earlier power-dependent PL studies of MSCs, we extended our analysis to PTM-passivated CsSnI$_3$ NCs. **Figure 5A** compares the PL at two different incident powers, with emission at higher power showing a blueshift and significant spectral broadening. The spectral broadening is documented in **Figure 5B**. As shown in greater detail in **Figure 5C**, the PL response has distinct regimes. At low excitation power densities (< 6 mW/μm²), the PL intensity remains largely flat, and the emission peak wavelength undergoes a modest blueshift from 930 nm to 920 nm. This behavior is characteristic of a trap-dominated regime, where photoexcited carriers are efficiently captured by deep or shallow defect states in the band gap, suppressing radiative recombination. The slight blueshift may result from the preferential recombination of higher-energy carriers before full thermalization but is insufficient to suggest hot-carrier effects. At approximately 6 mW/μm², a photo-brightening threshold is reached, where the trap states become saturated, and radiative recombination begins to dominate. Beyond this threshold, the PL intensity increases sharply, by nearly an order of magnitude, and the emission peak exhibits a more pronounced blueshift. This spectral shift reflects the presence of hot-carrier recombination, as carriers begin to recombine before reaching the band edge. Further, the PL intensity increasing with excitation power, coupled with the spectrum developing a high-energy exponential tail accompanied by a small blueshift of the peak, is behavior inconsistent with simple local lattice heating and characteristic of a nonequilibrium hot-carrier distribution. To quantitatively assess hot-carrier behavior, we analyzed the high-energy tail $I(E)$ of the PL spectrum using a Boltzmann-type exponential function of the form $I(E) \propto \alpha(E)e^{-E/k_B T_C}$ where α(E) is the absorption coefficient and $T_C$ is the effective carrier temperature.[32] As shown in the inset to **Figure 5A**, these fitting yields $T_C$ values ranging from 300 K to 1900 K as excitation power increases, confirmed by choosing different high-energy fitting windows, which show variations ~ 10%, indicated by error bars. These effective carrier temperatures $T_C$ are a standard diagnostic of hot-carrier populations in perovskite systems, and values in this range are consistent with hot-carrier photoluminescence (HCPL) observed in other HCPL-active lead-free and lead-based perovskites.[33-38]

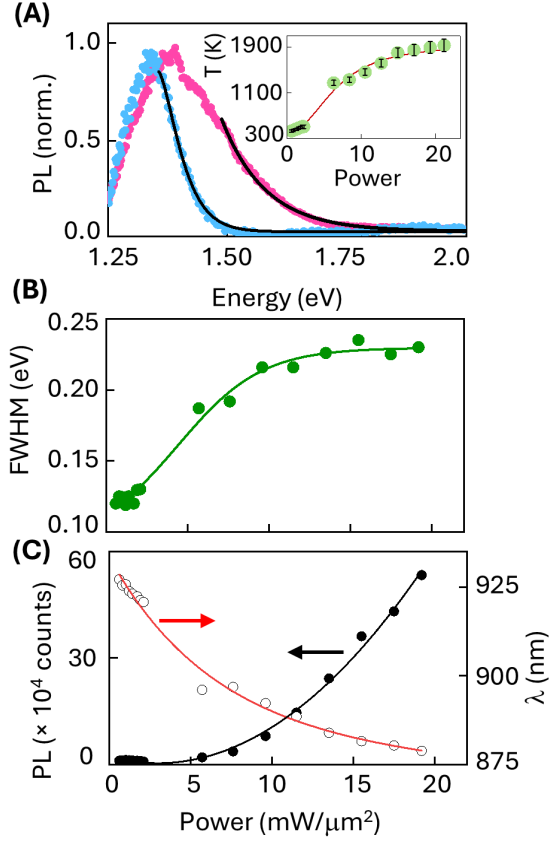

**Figure 5** (A) PL spectra at different excitation power. Lines on the higher energy tails are fits to $I_{PL} \propto \alpha(E)e^{-E/k_B T_C}$ and the carrier temperature $T_C$ extracted plotted in the inset as a function of excitation power. Error bars indicate the statistical variation acquired from extracting $T_C$ across multiple energy windows. (B) FWHM and (C) PL intensity (solid circles) and wavelength (open circles) varying with excitation power.

These results contrast the power-dependent PL behavior observed in the MSCs (**Figure 3A**), where redshift dominated and no hot-carrier effects were detected, as discussed in Section 2.3 The discrepancy likely arises from differences in carrier cooling dynamics between the two morphologies. In the MSCs, large thermal diffusion volumes, efficient phonon coupling, and defect-free diffusion pathways facilitate rapid carrier cooling. In contrast, the NCs, despite being PTM-passivated, have reduced thermal conduction pathways and possibly weaker carrier-phonon coupling, conditions which favor a transient hot-carrier population under strong excitation. Together, the emergence of a blueshift and a measurable increase in carrier temperature beyond the trap-saturation threshold provides strong evidence for hot-carrier recombination in PTM-passivated NCs.

### 2.6. Photoinduced Degradation and Reversible Trap Dynamics

To evaluate long-term photostability and trap behavior in PTM-passivated NCs, we tracked PL over multiple days under varying illumination conditions. Under continuous 532 nm excitation (**Figure 6A**), the PL intensity increases steadily throughout Day 1, reflecting a photo-brightening process. Concurrently, the emission peak redshifts (**Figure 6B**), which suggests that prolonged illumination induces reconfiguration of shallow trap states or local structural ordering, enhancing radiative efficiency. On Day 2, the sample was stored in the dark. Upon resuming illumination on Day 3, the PL intensity rises further, but the emission peak initially appears blueshifted relative to the endpoint of Day 1. This partial recovery indicates that some photoinduced lattice distortions or defect states anneal in the absence of light and re-emerge under renewed excitation. The data point to a reversible, dynamic defect landscape influenced by light exposure.

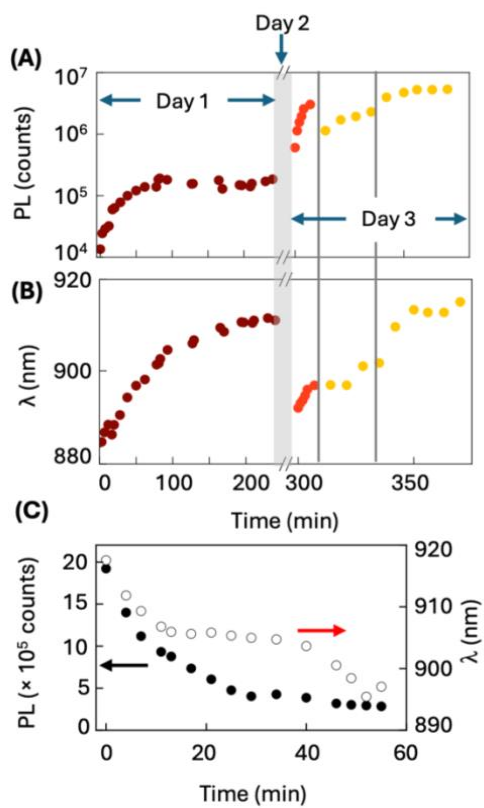

**Figure 6** (A) PL intensity and (B) wavelength of NC-PTM monitored under continuous long-term excitation. The shaded region indicates no photo exposure or measurements, and region enclosed between vertical lines indicate time over which measurements are performed without continuous excitation. (C) PL intensity (solid circles) and wavelength (open circles) of NC-PTM synthesized with 12% higher PTM concentration under continuous excitation (RH = 57%).

To assess the influence of PTM concentration, we also examined an NC batch prepared with 12% higher PTM loading. Unlike the prior samples, which exhibit stable emission, these display markedly different behavior (**Figure 6C**). Under identical excitation, these samples exhibit monotonic PL quenching and a progressive blueshift in the emission peak. This response likely stems from over-passivation disrupting lattice stability or facilitating $Sn^{2+}$ oxidation,[15] which leads to irreversible photobleaching. These results highlight a delicate balance in PTM loading; while moderate amounts stabilize and enable reversible defect dynamics, excessive PTM compromises long-term emissive integrity.

## 3. Conclusions

This study reveals a multi-dimensional picture of phase-stable, PTM-passivated $CsSnI_3$ microscale single crystals and nanocrystals, demonstrating how structure, size, and excitation conditions cooperatively influence optical and vibrational behavior. The key novel finding is the discovery of a two-stage photo-brightening mechanism in PTM-treated microscale crystals, correlated with the reversible sharpening of the $Cs^+$ rattling Raman mode. This coupling between vibrational coherence and radiative enhancement under light exposure is rare in lead-free perovskites and establishes a new framework for tuning optoelectronic properties via lattice dynamics.

By leveraging PL and Raman spectroscopy across spatial, temporal, and power domains, we differentiated between mechanisms governing carrier recombination. Microscale crystals display redshifts consistent with many-body bandgap renormalization, while nanocrystals exhibit blueshifting and hot-carrier photoluminescence, confirmed by PL tail fitting with extracted carrier temperatures up to 1900 K. The structural differences between morphologies further explain these dynamics: mechanically induced lattice reconstruction in ~150 nm nanocrystals leads to narrower, redshifted emission and increased susceptibility to reversible photoactivated trap formation. PTM remains effective in stabilizing both morphologies, mitigating degradation while permitting dynamic photophysical behavior to be probed. Consistent with this, the unchanged XRD patterns (*Supporting Information, Fig. S1*) and sustained NIR photoluminescence indicate that PTM-treated crystals do not undergo the $Sn^{2+} \rightarrow Sn^{4+}$ oxidation-driven transformation to $Cs_2SnI_6$. Together, these observations support the conclusion that PTM suppresses the $Sn^{2+}$ oxidation pathway under both illumination and ambient exposure.

Unlike previously studied Sn-based films or mixed-cation alloys, PTM-passivated $CsSnI_3$ microscale single crystals and nanocrystals exhibit morphology-tunable carrier cooling behavior, enabling direct comparison of phonon-dominated and hot-carrier regimes within a single material system. These findings provide fundamental insights into the relationship between lattice order, defect dynamics, and emission behavior in tin halide perovskites. Consistent with prior reports showing that $CsSnI_3$ thin films exhibit hundreds-of-picoseconds excitonic lifetimes while $CsSnI_3$ nanocrystals display few nanosecond decay components sensitive to surface chemistry and phase structure,[39,40] future work will leverage time-resolved photoluminescence to directly correlate these known morphology-dependent carrier lifetimes with the contrasting hot-carrier and bandgap-renormalization responses observed here. Additional opportunities include time-resolved vibrational spectroscopy and in situ oxidation-state tracking to better understand the mechanisms of trap migration and lattice reconfiguration. Another natural next step will be to perform temperature-dependent Raman and PL measurements to resolve how lattice anharmonicity, carrier–phonon coupling, and defect energetics govern the emergence and reversibility of the 40.8 cm$^{-1}$ mode. Such measurements will allow direct mapping of the vibrational free-energy landscape and will provide a more complete mechanistic picture of the light-modulated lattice coherence observed here. Extending this approach to related systems could yield general design rules for dynamically reconfigurable, lead-free NIR optoelectronics.

Together, these results establish PTM-passivated $CsSnI_3$ as a structurally dynamic, morphology-sensitive system that supports reversible optical modulation, vibrational coherence, and tunable carrier relaxation behavior across spatial and excitation regimes. This integrated picture provides new design principles for stable, responsive, and efficient lead-free perovskite optoelectronics.

## 4. Experimental Methods

*Materials:* Premium glass substrates purchased in Fisher Scientific. Isopropyl Alcohol ≥ 99.9%, CsI AnhydroBeads 99.999%, $SnI_2$ AnhydroBeads 99.99%, Phthalimide (PTM) ≥ 99%, N,N Dimethylformamide (DMF) 99.8%, and Acetonitrile (ACN) ≥ 99.9% in SigmaAldrich. Nitrogen Gas Ultra High Purity Grade in Airgas. Razor blades Surgical Carnon Steel Grade in VWR.

*Synthesis of CsSnI$_3$ and CsSnI3-PTM Perovskite Precursor Solutions*: 0.2054g of CsI and 0.2946g SnI$_2$ was dissolved in 1.8ml DMF and 1.2 ml ACN solution in a low humidity (16% - 17%) nitrogen filled glove box. The solution was left to magnetic stirring for 4 hours at approximately 1150 rpm in room temperature. The solution was then filtered through a 0.45μm syringe filter to yield the CsSnI$_3$ perovskite precursor solution.

CsSnI$_3$-PTM precursor solution was made by incorporating 1 mg of PTM (147.13 g/mol, Sigma-Aldrich) to 3 ml of unfiltered 0.267 M of CsSnI$_3$ solution, which yields PTM:CsSnI3 ratio of 0.0085. PTM was dissolved via magnetic stirring at approximately 1150 rpm for 5 minutes. The solution then followed the previously incorporated syringe filtration to yield the CsSnI$_3$-PTM precursor solution. The higher PTM-NC samples used PTM:CsSnI$_3$ ratio of 0.0095, 12% higher.

*Microscale Single Crystals (MSCs) and Nanocrystals (NCs) Fabrication*: First, 1.3 x 1.3 cm substrates cleaned with alcohol, dried with nitrogen gas and UV Ozone treated for 5 minutes. 2.5uL of CsSnI$_3$/CsSnI$_3$-PTM solution was deposited on the treated substrates and spin coated at 2000 rpm for 30 seconds. The coated substrate was annealed at 370 °C for 20 seconds to yield a deposition of MSCs.

MSC coated substrates were stripped with a razor and ground between cleaned substrates to yield a NC CsSnI$_3$/CsSnI$_3$-PTM black powder.

*Structural Characterization and Imaging*: X-Ray Diffraction (PANalytical X'Pert PRO Theta/Theta Powder X-ray diffraction System) was used to characterize the structural properties of CsSnI$_3$ and CsSnI$_3$-PTM MSCs. Confocal microscopy (LSM 880) was used to take absorption/emission and reflection images of the MSCs prepared on glass substrates. Samples were excited using a 532 nm laser. Reflection was filtered 400 nm–715 nm. Absorption/emission was characterized as black, absorption of 400–715 nm and emission > 715 nm. Scanning Electron Microscopy (Zeiss Gemini SEM 500) was used to characterize the size of CsSnI$_3$-PTM NCs.

*Optical Characterization*: UV-Vis absorption spectroscopy (Perkin Elmer Lambda 35 UV/Vis spectrophotometer) was used to measure absorbance of CsSnI$_3$ and CsSnI$_3$-PTM MSCs. Raman Spectroscopy, and PL spectroscopy were done using a 532 nm laser (ONDAX THz-Raman) as the excitation source with an excitation spot size of 1 μm$^2$. PL was characterized using a spectrometer (iHR 550) which utilized a 600 lines mm$^{-1}$ grating to disperse the collected light into a CCD device. Continuous-illumination stability measurements (Figs. 3B and 6A–C) were performed using a 532 nm excitation source at an incident power density of 0.2 mW/μm$^2$

**Supporting Information**

Supporting Information is available from the Wiley Online Library or from the author.

**Acknowledgements**

The authors acknowledge funding from NSF awards HRD-1547848 and DGE-2125510, the Sierra Nevada Research Institute (SNRI) Labor and Automation in California Agriculture (LACA) Graduate Summer Fellowship, and the Department of Defense through the National Defense Science & Engineering Graduate (NDSEG) Fellowship Program. XRD, SEM, and Confocal Microscopy experiments were performed at Imaging and Microscopy Facility (IMF) at UC Merced.

Received: ((will be filled in by the editorial staff))
Revised: ((will be filled in by the editorial staff))
Published online: ((will be filled in by the editorial staff))

# Light-Induced Lattice Coherence and Emission Enhancement in PTM-Passivated CsSnI$_3$ Perovskites

*Thomas Adams, Bruce Barrios, Michael Ziegenfus, Michael Scheibner, Hui Cai, and Sayantani Ghosh\**

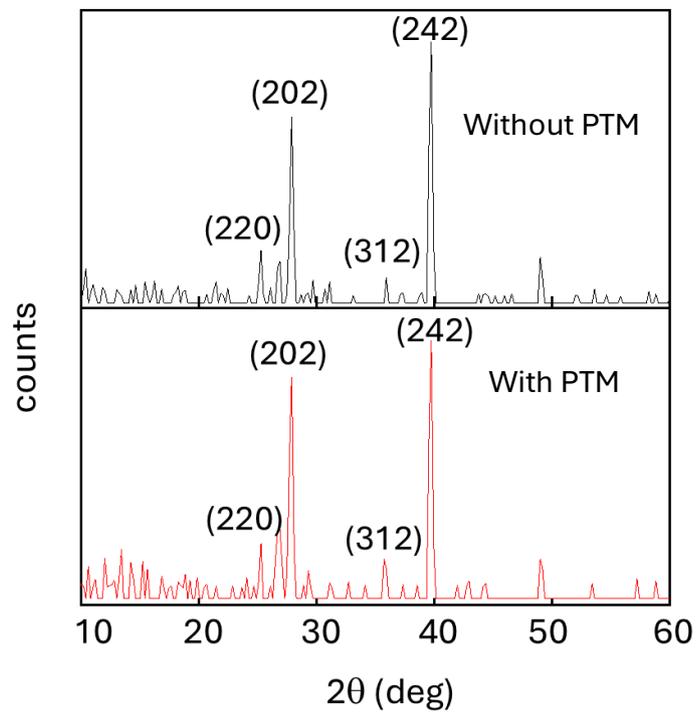

**Figure S1.** Untreated (top) and PTM-treated (bottom) MSCs show peak positions characteristic of the orthorhombic phase, with no detectable shift in lattice parameters. The absence of Cs$_2$SnI$_6$ reflections (13°, 28°) indicates no Sn$^{4+}$-driven phase conversion in PTM-treated samples

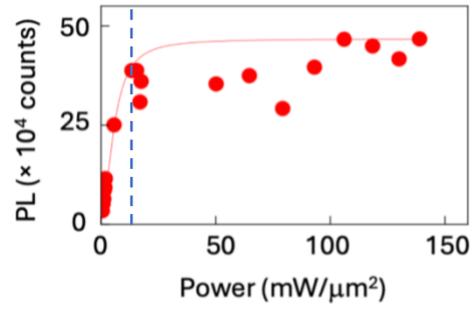

**Figure S2.** PL intensity as a function of incident power for MSCs. The dashed line denotes the incident power above which the emission starts redshifting, shown in Figure 3A. PL initially rises super linearly as more carriers are injected and recombine radiatively, but then saturates. It is compatible with the BGR onset, since redshifted emission often lags the PL rise due to the need for sufficient carrier density to induce screening.

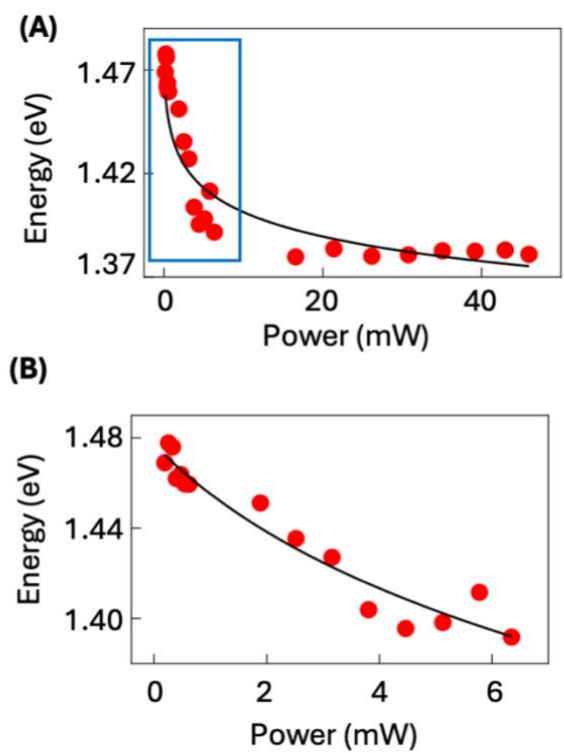

**Figure S3.** BGR fitted to emission energy as a function of incident power over (A) the entire power range and (B) low power range restricted to < 10 mW where the most significant redshift is observed. When fitted across the entire scale in (A) the fit returns $\Delta E$ = 0.017 eV, which is far lower than the observed redshift. The fit in (B) returns $\Delta E$ = 0.07 eV, an accurate measure of experimental result.

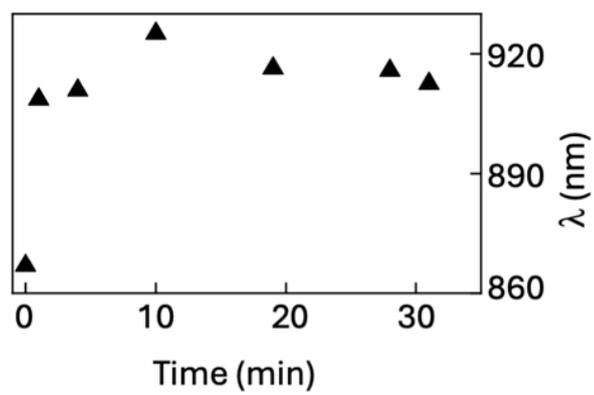

**Figure S4.** Emission wavelength of MSCs as a function of time during continuous illumination.

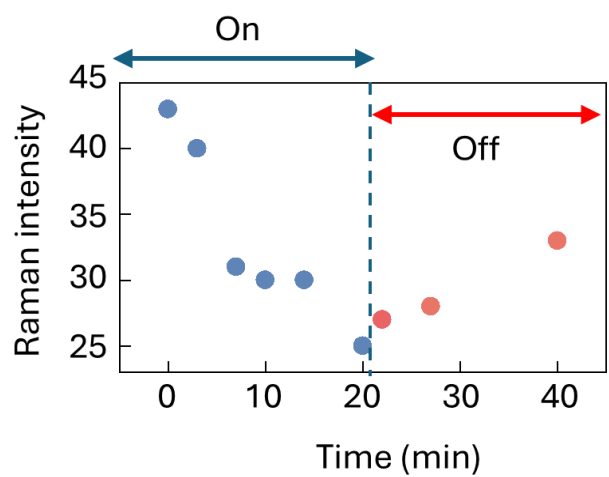

**Figure S5.** Intensity variation of the 40.8 cm$^{-1}$ Raman peak over time, as the continuous illumination is on, and then turned off, allowing the peak to recover.

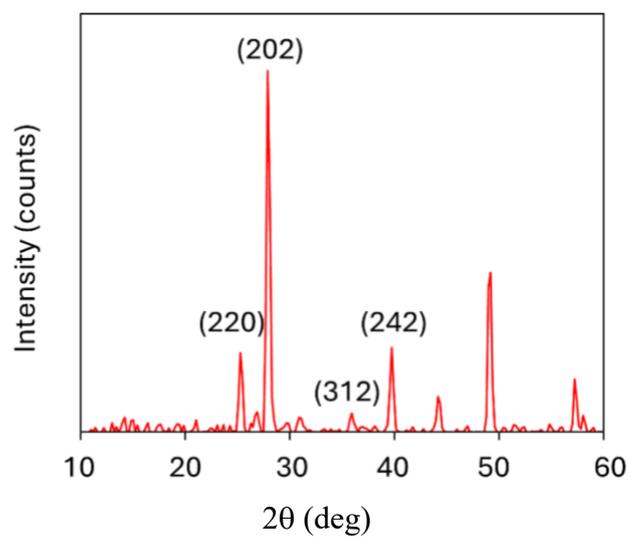

**Figure S6.** XRD of NCs confirms B-γ-CsSnI$_3$ phase with preferred 202 orientation and reduced 242 peak.